# Thermodynamics of markets

S.A. Rashkovskiy

*Ishlinsky Institute for Problems in Mechanics of the Russian Academy of Sciences, Vernadskogo Ave., 101/1 Moscow, 119526, Russia*

*Tomsk State University, 36 Lenina Avenue, Tomsk, 634050, Russia*

*E-mail: rash@ipmnet.ru, Tel. +7 9060318854*

**Abstract.** We consider the thermodynamic approach to the description of economic systems and processes. The first and second laws of thermodynamics as applied to economic systems are derived and analyzed. It is shown that there is a deep analogy between the parameters of thermodynamic and economic systems (markets); in particular, each thermodynamic parameter can be associated with a certain economic parameter or indicator. The economic meaning of such primordially thermodynamic concepts as pressure, volume, internal energy, heat, etc. has been established. The thermostatistics of the market is considered. It is shown that, as in conventional thermostatistics, many market parameters, such as price of goods, quantity of goods, etc., as well as their fluctuations can be calculated formally using the partition function of an economic system.

**Keywords:** econophysics, economic system; first and second laws; thermodynamic parameters of the economic system; thermostatistics of the economic system.

## 1. Introduction

Economics, as a theory [1], is fundamentally different from theories in physics, primarily in terms of its structure and principles of construction [1].

When constructing any theory, the most consistent is the axiomatic approach, when there is a small number of postulates (basic laws) that are a generalization of the results of observations, while the rest of the laws are a consequence of these postulates, and can be derived analytically (or logically) within the framework of this theory and its methods. This ideology is most fully implemented in physics. An example is thermodynamics, in which all laws and consequences of thermodynamics are derived on the basis of three basic laws.

Thermodynamics is undoubtedly one of the most successful physical theories. It arose and developed mainly as a phenomenological theory that generalizes experimentally observed facts and laws. Thermodynamics has provided a unified approach to describing a wide class of processes associated with temperature changes in a variety of physical systems. It seems that thermodynamic methods can be used not only in relation to physical systems, but also to systems outside physics, consisting of a large number of interacting elements. In such systems, specific mechanisms of interaction of elements become secondary, while certain collective properties come to the fore, which do not depend on the nature of the system, and should be the same for both physical and non-physical systems. It is these collective properties that determine the behavior of such systems as a whole.

These non-physical systems include social and economic systems, consisting of a large number of interacting actors (elements).

This paper considers a thermodynamic approach to describing economic systems and processes.

It should be noted that thermodynamic terminology is now widely used when it is necessary to characterize various economic processes. This includes, for example, terms such as pressure on markets, overheating of the market (economy), etc. These and other terms are used in economics intuitively, without any physical justification and without connection with the corresponding physical concepts.

In this work, we show that such a connection really exists and thermodynamic terms in economics have a deep physical meaning. In particular, we will establish the economic meaning of such primordially thermodynamic concepts as pressure, volume, internal energy, heat, etc.

The methods of thermodynamics and thermostatistics as applied to the description of economic systems and processes were considered in papers [2-14]. So in works [2,8-10,14], analogs of the first and second laws of thermodynamics are introduced and analyzed in relation to economic systems. The economic analogue of the Carnot cycle is considered in works [2,6-10]. In works [2-14], a new (for economics) quantitative indicator, temperature, is introduced and its economic meaning is discussed. In works [2-5,8,11-14], various economic systems and processes are analyzed by methods of statistical physics and the connection of economic temperature with probability distributions of wealth and income has been shown.

At the same time, it should be noted that in the cited works, thermodynamic methods and thermodynamic terminology were often used intuitively and formally, at the level of external analogies, without strict theoretical justification. In this work, we try to construct and substantiate economic thermodynamics as a phenomenological theory, by analogy with how it is done in physics.

To show the deep connection of the theory under consideration with thermodynamics, we deliberately use thermodynamic terminology when describing economic systems and processes, while indicating the corresponding economic analogs of the thermodynamic terms and concepts used.

## 2. First law

We consider the simplest economic system, on the example of which we will analyze the thermodynamics of markets. In the future, we will try to complicate the system under consideration, bringing it closer to real economic systems.

Let there be a certain group of market actors (people, firms, etc.) which can exchange by money, goods, etc. We will call this group an economic system or simply a system (market). Market actors are elements of this system. We will assume that the system under consideration has $N \gg 1$ elements. Each $i$-th element of the system has a certain amount of money $\varepsilon_i, i = 1 \ldots N$. Accordingly, the total amount of money that is in the system is

$$E = \sum_{i=1}^{N} \varepsilon_i \tag{1}$$

We assume that there be certain goods that is needed by all elements of the system, and they are ready to buy it. This goods exists both in the system itself (it is possessed by the elements of the system) and outside the system (that is, it belongs to elements of other similar systems). We denote by $V$ the amount of goods that the elements of the system under consideration already have. This amount of goods is distributed among the elements of the system, generally speaking, non-uniformly and randomly. The goods can be bought and sold. Elements of the system under consideration can purchase goods from another system (from elements of another system), and can sell it to another system (to elements of another system). We assume that the mean price at which a given good is bought and sold is $p$. Thus the mean price $p$ at which the elements of a given system buy goods at a given moment from other systems is equal to the mean price at which elements of this system sell this goods at the same moment to other systems. From this point of view, the market under consideration is primitive. Note that $p$ is precisely the mean price of a unit of goods (one piece, one kilogram, etc.) in a given system, while each specific element of the system can buy or sell this goods at its own, generally speaking, random price.

Suppose that as a result of the purchase and sale of goods, its quantity in the system has changed by $dV$. If $dV > 0$, it means that the system has bought an additional quantity of goods. In doing so, it spent part of the money it had. As a result of this operation, the amount of money in the system decreased by the amount

$$dE = -pdV \tag{2}$$

If the system, on the contrary, has sold to other systems a part of the goods it has, then $dV < 0$ and the amount of money in the system has increased by an amount determined by the same relation (2).

The considered method of changing the amount of money in the system is associated with the purchase and sale of goods.

There is another way to change the amount of money in the system, which is not related to the sale and purchase of goods, when the elements of this system directly exchange money (i.e., receive or give money) with elements of some external (in relation to this) system. This can be, for example, direct investments, dividend payments, cash gifts, donations, taxes, subsidies, loans, loan payments, etc.

Let the amount of money which transferred between the system under consideration (all elements of this system) and external systems without changing the amount of goods $V$ is equal to $\delta Q$. We assume that if the system received an additional amount of money in this way, then $\delta Q > 0$, while if it gave money, then $\delta Q < 0$.

During this process (at $V = const$) the amount of money that is in the system will change by the amount

$$dE = \delta Q \qquad (3)$$

So far, we have considered a system with a constant number of elements. However, elements (market actors) can move from the system (market) under consideration to other systems, and, conversely, can come to the system under consideration from other systems. In this case, the number of elements of the system $N$ is variable. Moving from system to system, the elements carry a certain amount of money with them. We denote $\mu$ the mean amount of money that one element carries with it, passing from one system to another. Note that when moving to another system, the element carries with it not only money, but also some "activity"; thus parameter $\mu$ characterizes also the change in the mean activity of elements in the system when the number of elements changes. Then, if in the course of a certain process the number of elements in the system changed by $dN$ due to the transition (migration) of elements, then the amount of money in the system changes by

$$dE = \mu dN \qquad (4)$$

Another process leading to a change in the number of elements in the system is possible. To consider this process, we clarify the concept of a system element. The element is the minimum structural unit of the system, which appears in various processes as a single whole. For example, the elements of the economic system can be at the same time, individuals, families, firms, companies, etc. It follows that elements can be simple or complex. Simple elements remain unchanged throughout the process under consideration. In particular, they cannot break down into simpler ones, but they can be combined into more complex elements. Obviously, in relation to the considered economic systems (markets), people themselves are always simple elements. They can unite in families, firms, companies, etc .; in turn, small firms and companies can merge into larger ones, etc. As a result, complex elements of the system are formed that participate in economic processes as a single whole, but at the same time they themselves consist of simpler elements. By analogy with physics and chemistry, the process of combining simpler elements into more complex ones will be called recombination. Under certain conditions, complex elements can break down into simpler ones. This process will be called dissociation by analogy with physics and chemistry. Note that the same element in some processes can be considered as simple, while in others as complex.

For example, if during a certain process taking place in the system, some firm or company does not split into smaller ones, then it should be considered as a simple element, otherwise it is a complex element. As in physical systems, in real economic systems, the processes of recombination and dissociation take place simultaneously: constantly one family, firms, companies, etc. are formed, while others are disintegrated. Note that these processes are similar to chemical reactions, for example, in gases. In the processes of dissociation-recombination, the number of elements of the system changes and, at the same time, this is accompanied by a change in the amount of money available in the system. This is due to the fact that in the processes of dissociation and recombination of elements of the economic system, part of the money can be spent or, on the contrary, released as a result of the decay or combination of elements. If the change in the number of elements in the system due to the dissociation-recombination of elements is equal to $dN$, then the total change in the amount of money in the system is also described by the relation (4), where $\mu$ is the change in the amount of money in the system when the number of system elements changes per unit due to dissociation-recombination of elements.

In the general case, the amount of money in the system can change simultaneously due to the sale and purchase of goods, due to the change in the number of system elements and due to the direct exchange of money between the elements of different systems, not associated with the sale and purchase of goods or with the exchange of elements.

Putting together relations (2)-(4), we obtain in the general case an equation describing the change in the amount of money in the system under consideration:

$$dE = \delta Q - pdV + \mu dN \qquad (5)$$

or

$$\delta Q = dE + pdV - \mu dN \qquad (6)$$

This equation in its form exactly coincides with the first law of thermodynamics. In the context of economic systems, we will also call it the first law.

Formal comparison of equation (6) with the first law of thermodynamics allows concluding that the amount of money $E$ in the economic system plays the role of internal energy, the amount of goods in the economic system $V$ plays the role of the volume of the system, the mean price of a unit of goods $p$ plays the role of pressure, the mean change in the amount of money in the system with a change in the number of its elements per unit (both due to migration and due to dissociation-recombination), $\mu$, plays the role of a chemical potential. As applied to economic systems, the parameter $\mu$ will be called the financial potential of the system [14].

Continuing the thermodynamic analogy, we come to the conclusion that the amount of money $\delta Q$, which is directly transferred from the elements of one system to the elements of another system, without changing the amount of goods and the number of elements of the system, plays the role of thermal energy or heat.

Note that the first law in the form (6) as applied to economic systems was considered in [2,7-10,13,14], but there it was introduced formally, and the meaning of the parameters entering it remained unclear. In this work, we derived the first law in the form (6), based on understandable economic considerations with a full understanding of what the parameters included in it mean.

As a result of this analysis, we come to an important conclusion: money in the economic system plays the same role as internal energy in a physical thermodynamic system. This analogy is not accidental; it has deep roots. Consider this using an ideal gas as an example. The energy of an ideal gas is made up of the energy of all the atoms and molecules that make up this gas. Similarly, the amount of money available in an ideal (primitive) market is equal to the sum of all money that the actors (elements) of this market have (1). For money, as well as for energy, the conservation law is fulfilled. Indeed, consider two market actors. Let them have, respectively, $\varepsilon_1$ and $\varepsilon_2$ money. These elements can interact with each other, for example, selling goods to each other, or simply by transferring money to each other (investing, borrowing, donation, etc.). Let, as a result of this interaction of the elements, the amount of money of each of them changed and became, respectively, $\varepsilon'_1$ and $\varepsilon'_2$. Obviously,

$$\varepsilon_1 + \varepsilon_2 = \varepsilon'_1 + \varepsilon'_2 + \delta\varepsilon \tag{7}$$

where $\delta\varepsilon$ is the amount of money that passed to third parties as a result of this interaction. The value $\delta\varepsilon$ describes possible losses (taxes, physical loss of money, etc.) or the acquisition of money as a result of a given financial transaction. In other words, like energy, money does not disappear anywhere and does not appear from anywhere; they can only move from one system (element) to another or from one type to another (for example, from one currency to another, from cash to non-cash, etc.). In any case, we can always draw up a financial balance that will converge absolutely. The role of money as energy in the economic system was discussed in [8,14].

It follows that, as in the thermodynamics of physical systems, the first law (5) and (6) expresses the law of conservation of money (energy) for economic systems (markets).

Taking into account this analogy, in the future, in relation to economic systems, for brevity, we will use the following concepts:

- internal energy of system $E$ - the amount of money available in the system (further we will clarify the concept of energy),

- volume of the system $V$ - the amount of goods in the system,

- pressure $p$ - mean price of goods in the system,

- financial potential $\mu$ - a change in the amount of money in the system when the number of its elements changes per unit (due to migration, dissociation, recombination, etc.),

- heat $\delta Q$ - the amount of money that the elements of one system directly transfer to the elements of another system without buying and selling goods and without changing the number of elements in the system.

The first law (5) shows that there are three ways to change the energy of an economic system:

(i) By changing the volume of the system. By analogy with thermodynamics, this method of changing energy will be called work. The work done by the system when its volume changes by $dV$ is equal to

$$\delta W = pdV \qquad (8)$$

(ii) By changing the number of system elements. This way of changing energy will be called financial work (4).

(iii) Without changing the volume of the system and the number of its elements due to the direct transfer of money between the elements (3). By analogy with thermodynamics, this method of changing energy will be called heat transfer or heat exchange.

There are no other ways to change the energy (the amount of money) of the economic system. It is assumed here that money can be emitted or withdrawn from circulation only by the Central Bank (in the USA it is the Federal Reserve), which does not belong to the systems under consideration. Note that the Central Bank (Federal Reserve) plays the same role in the economic system as the thermostat (heat reservoir) plays in conventional thermodynamics.

Depending on the nature of the interaction of the system under consideration with external systems, various types of economic systems can be considered:

- open system - a system (market) that can exchange energy and elements with other systems;

- an isolated system - a system (market) that does not exchange energy or elements with other systems;

- closed system - a system (market) that does not exchange elements with other systems, but can exchange energy.

- an adiabatically isolated or adiabatic system - a closed system (market) that can do work, but cannot exchange energy in the form of heat with other systems.

## 3. Economic system temperature and equations of state

The amount of money that the system (market) could potentially earn by selling all the goods it has at the mean price $p$ is $pV$. In reality, the energy $pV$ is only the "potential" energy of the system, since if all elements of the system (market actors) simultaneously want to sell the entire goods, then the price for it will fall and the system will not receive the expected amount of money. However, $pV$ is some energy characteristic of the system.

We introduce the "potential" energy of the system per one element of the system

$$k_0 T = \frac{pV}{N} \qquad (9)$$

The new parameter $T$ introduced in this way will be called the economic temperature or simply the temperature of the system (market) under consideration. Here, $k_0 > 0$ is a constant scale factor depending on the choice of the temperature unit. Temperature $T$ is some additional characteristic of the economic system (market), along with energy, volume and pressure. As in thermodynamics, the temperature determined by relation (9) should be considered as a special parameter that, along with other parameters (price of goods, volume of goods, number of elements, etc.) characterizes the system in a state of equilibrium. The temperature determined by relation (9) is always positive.

The concept of economic temperature as an additional parameter that characterizes the economic system was discussed in [2-14]. Further, we will give a general definition of economic temperature and clarify its meaning.

The relation (9) can be rewritten as

$$pV = k_0 NT \qquad (10)$$

Relation (10) can be considered as an equation connecting pressure, volume and temperature of the system (market) under consideration. It indicates that the three parameters $T, p, V$, characterizing the system, are not independent, but are related by expression (10), which we will call the equation of state of the economic system. The equation of state (10) formally coincides with the Mendeleev-Clapeyron equation of state for an ideal gas. In [14], the equation of state (10) was obtained as a limiting case, within the framework of the statistical analysis of the economic system. By analogy with thermodynamics, economic systems obeying the equation of state (10) will be called ideal (primitive) markets.

For real financial and economic systems, one can also introduce the temperature and the equation of state

$$p = f(T, V, N) \qquad (11)$$

One can expect that different systems have different equations of state, generally speaking, differing from (10).

Equations of state, similar to (11), which relate the pressure, volume and temperature of the system, by analogy with thermodynamics, will be called thermal equations of state, since they are, in fact, the definition of economic temperature.

Energy, volume and pressure fully characterize the equilibrium state of the considered economic system. At the same time, they are not independent parameters: when the energy of the system (the amount of money in the system) and/or its volume (the amount of goods in the system) change, the pressure (the mean price of the goods) changes simultaneously.

This means that there is some equation

$$\varphi(E, p, V, N) = 0 \qquad (12)$$

which establishes a relation between the parameters $E, p, V, N$ for an equilibrium system.

Substituting the pressure $p$ from (11) into equation (12), we obtain

$$E = E(T, V, N) \qquad (13)$$

Relation (13) shows that the energy of an economic system in an equilibrium state is completely determined by its temperature and volume. For this reason, relation (13) can be called the energy (financial) equation of state of the economic system.

By analogy with the thermodynamics of physical systems, for the financial potential $\mu$ of an equilibrium system there must be a dependence

$$\mu = \mu(T, V, N) \qquad (14)$$

which plays the role of the third equation of state of the economic system.

The parameters that describe the system can be divided into two groups. The first group includes parameters proportional to the number of elements in the system. As usual, such parameters will be called extensive. The second group includes parameters that, other things being equal, are not proportional to the number of system elements. Such parameters will be called intense. To understand whether a particular parameter of an economic system is intensive or extensive, it is enough to mentally divide the system which is in equilibrium into two equal parts and see what the parameters of each of the parts are equal to. If a parameter of each of the parts of the system is equal to the same parameter for the entire system as a whole, then this parameter is intense. If the parameter of each of the parts is equal to half of the same parameter for the entire system as a whole, then this parameter is extensive. The ratio of two extensive parameters is an intensive parameter. Obviously, the number of elements $N$, energy $E$ and volume $V$ of the system are extensive parameters, while pressure (mean price of a unit of goods), economic temperature $T$, financial potential $\mu$, density $\rho = N/V$ and specific volume $V/N = 1/\rho$ (the amount of goods per element) of the system are intensive parameters. From the definition of economic heat, it follows that $\delta Q$ is also an extensive parameter: the thermal energy (in the economic sense) received by the system is redistributed among all elements of the system so that the thermal energy (amount of money) received on average by each part of the system is proportional to the number of elements in these parts.

Obviously, intensive parameters can only depend on intensive parameters or on the ratio of extensive parameters. Thus, the equations of state (11), (13), and (14) can be rewritten as

$$pV = Nf_p(T, V/N) \qquad (15)$$

$$E = Nf_E(T, V/N) \qquad (16)$$

$$\mu = f_\mu(T, V/N) \qquad (17)$$

Taking into account (15), equation (17) can be written in the form

$$\mu = \mu(T, p) \qquad (18)$$

which is an alternative record of the equation of state (14) of the economic system.

## 4. Entropy of the economic system and second law

Under certain conditions, the economic system (market) is asymptotically stable: the system returns to its original state if it was removed from it by a short-term external influence (disturbance). This state of the economic system will be called the equilibrium state. So the equilibrium state of the economic system is its attractor.

If the system has become unstable for some reason, it will be brought out of this state by disturbances that always exist, and will change until it finds a new stable state. Most of the time, any system is in a stable (equilibrium) state. Otherwise, we would always observe unstable and unpredictable behavior of the market (system), on which it would be impossible to work.

The stability of a particular system can be easily verified directly: if the acting external disturbances (which always take place) do not lead to a significant and rapid change in the state of the system, then it is stable.

The fact that the system is in a stable state indicates that internal processes are taking place in it that keep the system in an equilibrium state, i.e. seeking to return it to this state.

Consider an isolated system which has an asymptotically stable (equilibrium) state. For such a system, we can introduce the Lyapunov function $S$, which, at $E = const$, $V = const$ and $N = const$, increases if the system is not in equilibrium, and remains constant if the system is in equilibrium:

$$dS \geq 0 \qquad (19)$$

where the equal sign corresponds to an equilibrium state, while the inequality sign corresponds to a nonequilibrium state, and indicates the tendency of the system to an equilibrium state. From condition (19), it follows that for $E = const, V = const$ and $N = const$, the economic system which is in equilibrium has the maximum value of the Lyapunov function $S$ in comparison with any of its other states.

Let the Lyapunov function $S$ in the equilibrium state take the value $S^0$.

Because the equilibrium state of the economic system (market) is completely determined by its energy and volume, we can write

$$S^0 = S^0(E, V, N) \tag{20}$$

According to definition (20), the parameter $S^0$ is a new parameter of the state of the system, along with $E, V$ and $N$.

It is known from economic theory [1] that when the supply (quantity of goods) changes, the price of the goods changes (the law of supply and demand). With regard to the system under consideration, this means that a change in the volume of the system (the quantity of goods), other things unchanged, leads to a change in the pressure (price of the goods) in the system. Similarly, a change in the energy of the system (the amount of money in the system), all other things being equal, leads to a change in the pressure and/or volume of the system.

Thus, by changing the volume of the system and/or exerting a thermal (in the economic sense) effect on it, one can change the state of the system and perform work (sell or buy goods).

We come to the concept of a thermodynamic process in an economic system, which is characterized by a change in the parameters of the system $E, p, V, N$ and $T$. Note that one should not confuse the thermodynamic process in the economic system, which is associated with changes in the system as a whole, with internal processes in the economic system associated with changes in the state of individual elements or groups of elements, which can occur at constant values of the parameters $E, p, V, N$ and $T$.

Here, as in the thermodynamics of physical systems, one can speak of equilibrium and nonequilibrium processes. An equilibrium process, as usual, is understood as a process in which the system at each moment of time is in one of the equilibrium states. Otherwise, the process will be nonequilibrium. Obviously, an equilibrium process is an idealization: no real process is actually equilibrium, however, very slow (quasistatic) processes can be approximately considered as equilibrium. Let the characteristic time of the change in the volume of the system or its energy be equal to $\tau_V$ (this time characterizes the rate of change of the system as a whole due to the external influence), while the characteristic relaxation time of the internal state of the system is equal to $\tau_p$ (this time characterizes the rate of adaptation of the system to a new state, to new conditions). The process can be considered quasistatic, and therefore equilibrium, if the condition

$$\tau_p/\tau_V \ll 1 \tag{21}$$

If condition (21) is not met, then the process is non-static and, accordingly, nonequilibrium.

The equilibrium process is obviously reversible: when the parameters of the system $E, p, V, N$ and $T$ change in the opposite direction, the system goes through the same equilibrium states as in the direct process, but in the back sequence. For example, if we slowly increase the quantity of goods $V$ in a primitive market (all other things being equal), then according to the law of supply and demand [1], the average price of this goods $p$ decreases. If we again slowly reduce the quantity of goods on the market to the original volume, then the price of goods in the primitive market will return to its previous value (this does not take into account such a phenomenon as inflation, which is associated with nonequilibrium processes).

Nonequilibrium processes are obviously irreversible.

For an elementary equilibrium process, taking into account (20), we obtain

$$dS^0 = \left(\frac{\partial S^0}{\partial E}\right)_{V,N} dE + \left(\frac{\partial S^0}{\partial V}\right)_{E,N} dV + \left(\frac{\partial S^0}{\partial N}\right)_{E,V} dN \tag{22}$$

We introduce the parameter $T$:

$$\frac{1}{T} = \left(\frac{\partial S^0}{\partial E}\right)_{V,N} \tag{23}$$

which we will call the absolute economic temperature or simply the temperature of the economic system.

Relationship (23) is a general definition of the economic temperature of the system, while relationship (9) is only a special case related to the ideal (primitive) market. Note, that the thermal equation of state (11) must include the economic temperature determined by the relation (23).

Substituting (5) into (22) taking into account (23), we obtain

$$TdS^0 = \delta Q + \left[T\left(\frac{\partial S^0}{\partial V}\right)_{E,N} - p\right]dV + \left[T\left(\frac{\partial S^0}{\partial N}\right)_{E,V} + \mu\right]dN \tag{24}$$

Parameters $V, N, S^0$ can be changed independently.

Note that the Lyapunov function $S$ is not uniquely defined: any monotonically increasing function of $S$ can also be considered as the Lyapunov function of this system. In the general case, consider the function

$$S' = F(S, E, V, N) \tag{25}$$

where $F(S, E, V, N)$ is a function monotonically increasing on $S$, i.e. $\left(\frac{\partial F}{\partial S}\right)_{E,V,N} > 0$, and arbitrarily depending on $E, V, N$.

Obviously, for an isolated system with constant parameters $E, V, N$, for which condition (19) is satisfied, the condition $dS' \geq 0$ is also satisfied. So function (25) can also be considered as a Lyapunov function.

In the equilibrium state of the system, function (25) takes the value

$$S^{0'}(E, V, N) = F(S^0(E, V, N), E, V, N) \tag{26}$$

This ambiguity allows imposing additional limitations on the Lyapunov function.

We choose as function (20) such that satisfies the conditions

$$T\left(\frac{\partial S^0}{\partial V}\right)_{E,N} = p \tag{27}$$

$$T\left(\frac{\partial S^0}{\partial N}\right)_{E,V} = -\mu \tag{28}$$

The Lyapunov function $S$ satisfying conditions (27) and (28) will be called the entropy of the economic system.

Then equation (24) takes the form

$$TdS^0 = \delta Q \tag{29}$$

or

$$dS^0 = \frac{\delta Q}{T} \tag{30}$$

Taking into account the first law (6), we rewrite (30) in the form

$$TdS^0 = dE + pdV - \mu dN \tag{31}$$

Relations (29), (30) and (31) will be called the second law for equilibrium processes in economic systems.

It shows that a change in the entropy of an economic system in an equilibrium process is associated only with heat exchange (in the economic sense): the entropy of an economic system in an equilibrium process can only be changed by supplying heat to the system or removing it from the system. From an economic point of view, this means that it is possible to change the entropy of a system in an equilibrium process only as a result of a direct exchange of money between the elements of a given system and elements of systems external to it. As in physics, one can give a statistical meaning of the entropy of an economic system [2-14].

Note that even under conditions (27) and (28), entropy (20) and temperature (23) are not uniquely determined. Indeed, by choosing another function (26), we simply redefine the absolute temperature $T' = T\left(\frac{\partial F}{\partial S^0}\right)^{-1}$ without changing the second law.

As indicated above, similar to thermodynamics, the process taking place in the economic system at $\delta Q = 0$, i.e. without direct exchange of money between the elements of a given system and elements of other systems, can be called adiabatic.

According to (30), the equilibrium adiabatic process in the economic system is isentropic ($S = const$).

Relations (27) and (28) lead to the condition

$$\left[\frac{\partial}{\partial N}\left(\frac{p}{T}\right)\right]_{E,V} = -\left[\frac{\partial}{\partial V}\left(\frac{\mu}{T}\right)\right]_{E,N} \quad (32)$$

which shows that equations of state (11), (13) and (14) are not independent.

Using the energy equation of state (13), we write (20) in the form

$$S^0 = S^0(T, V, N) \quad (33)$$

From the definition (30), it follows that entropy is an extensive parameter, and, therefore, we can write

$$S^0 = N f_S^0(E/N, V/N) \quad (34)$$

or

$$S^0 = N f_S(T, V/N) \quad (35)$$

Consider an economic system (market) in equilibrium. We consider this system to be homogeneous. This means that if we mentally divide the system under consideration into two parts (into two subsystems), then the intensive parameters ($p, T, \mu$, etc.) for each part will be the same and coincide with the corresponding parameters of the entire system, while the sum of extensive parameters ($E, V, N, S^0$, etc.) of each subsystem will be equal to those for the entire system. These subsystems can be viewed as separate systems that interact with each other (being part of a large system) and at the same time can interact with external systems. By condition, the considered subsystems (parts of a large system) are in equilibrium. Thus, we come to the conclusion that a sufficient condition for the equilibrium of two systems (markets) is the equality of all their intensive parameters.

Consider now the case when there are two interacting economic systems that are in equilibrium states, but have different economic temperatures $T_1$ and $T_2$. Let the total amount of heat (in the economic sense) that system 1 receives in the course of an elementary process is equal to $\delta Q_{\Sigma 1}$. The amount of heat that system 2 receives during this process is equal to $\delta Q_{\Sigma 2}$. The corresponding changes in the entropy of these systems are described by relation (30) and are equal to

$$dS_1^0 = \frac{\delta Q_{\Sigma 1}}{T_1}, dS_2^0 = \frac{\delta Q_{\Sigma 2}}{T_2} \quad (36)$$

Consider a combined system consisting of two considered systems, which can also interact with each other and with external systems. Since the entropy is an extensive parameter, the entropy $S$ of the combined system is

$$S = S_1^0 + S_2^0 \quad (37)$$

Accordingly, the change in the entropy of the combined system is
$$dS = dS_1^0 + dS_2^0 \qquad (38)$$
Taking into account (36), we rewrite (38) in the form
$$dS = \frac{\delta Q_{\Sigma 1}}{T_1} + \frac{\delta Q_{\Sigma 2}}{T_2} \qquad (39)$$
The total amount of heat $\delta Q_{\Sigma 1}$ that system 1 receives is the sum of the heat $\delta Q_1$ that this system receives from external systems and the heat $\delta Q_{12}$ that this system receives from system 2:
$$\delta Q_{\Sigma 1} = \delta Q_1 + \delta Q_{12} \qquad (40)$$
Similarly,
$$\delta Q_{\Sigma 2} = \delta Q_2 + \delta Q_{21} \qquad (41)$$
where $\delta Q_2$ is the amount of heat that system 2 receives from external systems; $\delta Q_{21}$ is the amount of heat that system 2 receives from system 1.

If heat (money) is transferred between the systems under consideration directly, then
$$\delta Q_{21} = -\delta Q_{12} \qquad (42)$$
Taking into account relations (40) - (42), equation (39) takes the form
$$dS = \left(\frac{1}{T_1} - \frac{1}{T_2}\right)\delta Q_{12} + \frac{\delta Q_1}{T_1} + \frac{\delta Q_2}{T_2} \qquad (43)$$
First, consider the case when the systems under consideration do not receive heat from external systems ($\delta Q_1 = \delta Q_2 = 0$), but can exchange heat with each other ($\delta Q_{12} \neq 0$). In this case, equation (39) takes the form
$$dS = \left(\frac{1}{T_1} - \frac{1}{T_2}\right)\delta Q_{12} \qquad (44)$$
Since the systems under consideration have different economic temperatures, they are not in equilibrium with each other; this means that the combined system is not in equilibrium. In this case, inequality (19) holds, i.e. the entropy of the combined system increases. Taking into account (44), we can write inequality (19) for the considered combined system in the form
$$\left(\frac{1}{T_1} - \frac{1}{T_2}\right)\delta Q_{12} \geq 0 \qquad (45)$$
where the equal sign corresponds to the equilibrium state, while the inequality sign corresponds to the nonequilibrium state.

Inequality (45) holds in two cases:
$$\text{either } (T_1 > T_2, \delta Q_{12} < 0) \text{ or } (T_2 > T_1, \delta Q_{12} > 0) \qquad (46)$$
From conditions (46), taking into account (42), it follows that heat (in the economic sense) spontaneously flows from a system (market) with a high economic temperature to a system (market) with a lower economic temperature.

Based on this result, even without knowing a meaning of the economic temperature (which we will discuss below), we can argue that a system from which money spontaneously (without doing work) flows into another system has a higher economic temperature than the system in which this money flows.

Now, consider the case when systems can exchange heat with external systems ($\delta Q_1 \neq 0$; $\delta Q_2 \neq 0$). In this case, the change in the entropy of the combined system is described by equation (43). Taking into account that the heat transfer between the systems satisfies inequality (45), from equation (43) we obtain

$$dS \geq \frac{\delta Q_1}{T_1} + \frac{\delta Q_2}{T_2} \qquad (47)$$

Inequality (47) can be generalized to an arbitrary nonequilibrium system consisting of $K$ interacting subsystems, each of which is in a near-equilibrium state:

$$dS \geq \sum_{k=1}^{K} \frac{\delta Q_k}{T_k} \qquad (48)$$

where $T_k$ is the temperature of $k$-th subsystem; $\delta Q_k$ is the amount of heat that the $k$-th subsystem received from external systems; the equal sign corresponds to an equilibrium process, while the inequality sign corresponds to a nonequilibrium one.

Relation (48) is the second law for nonequilibrium processes in economic systems.

In the particular case, when the temperatures of the subsystems of the system under consideration differ little, i.e.

$$T_1 \approx T_2 \approx \cdots \approx T_K \approx T \qquad (49)$$

inequality (48) can be written as

$$dS \geq \frac{\delta Q}{T} \qquad (50)$$

where $\delta Q = \sum_{k=1}^{K} \delta Q_k$ is the total amount of heat that the combined system received from external systems.

Inequality (50) coincides with the traditional form of the second law of thermodynamics for nonequilibrium processes, however, as we see, it is valid only for weakly nonequilibrium systems. In the general case, the second law for nonequilibrium processes has the form (48).

## 5. Thermostatistics of markets

### 5.1. Grand canonical distribution

The thermodynamics of physical systems finds its justification in statistical physics. It can be expected that economic thermodynamics can also be substantiated within the framework of the thermostatistical analysis of economic systems.

The statistical mechanics of money was considered in works [3-8, 11-13]. An interesting analogy has been drawn between the economic system and a system of ideal gas where particles and their energies are modeled as agents and their wealth. Collision between particles is similar to trading between agents where energy or wealth is neither created nor destroyed; it is just redistributed between agents. As pointed out in [3-8], such a process obviously generates Gibbs-like distributions of wealth observed for the majority of the population. In these works, only money was considered as wealth, while non-monetary assets (including goods) that could be possessed by the elements of the system were not taken into account.

As shown above, in the thermodynamic description of the market, it is necessary to take into account the exchange between the elements of the system not only in the form of money, but also in the form of a commodity.

Consider the thermostatistics of markets using the statistical physics approach [15].

There is a system (market) in equilibrium. Let the energy of this system (the amount of money) be equal to $E$, the volume of the system (the amount of goods) equal to $V$, and the number of elements equal to $N$. We mentally divide this system into two parts. This division can be implemented in several ways. For example, we can split the system so that each element belongs to only one of the subsystems, or we can split it so that each unit of goods will belong to only one of the subsystems. In cases where each element of the system owns only one unit of goods, or each unit of goods belongs to only one element, these two ways of dividing the system are equivalent. However, if an element simultaneously owns several units of goods, then with the first method of division, the element is always in a certain subsystem, while different units of goods that it owns may belong to different subsystems. If we choose the second method of division, then such elements can refer simultaneously to both subsystems. In this case, a certain connection (interaction) exists between the two subsystems, and they will no longer be strictly independent. Another situation arises if certain goods have several owners, i.e. several people can simultaneously dispose of these goods (for example, members of the same family). Then, in the first variant of division, different owners of one goods may end up in different subsystems, while the goods that they simultaneously own will simultaneously belong to both subsystems (i.e., the elements will be interconnected by this goods). If we choose the second variant of division, then such goods belong to one of these subsystems, while its owners (elements) may end up in different subsystems. In this case, a certain connection (interaction) will also arise between the systems, and they will no longer be completely independent. In this work, we choose the first way to divide the system into two subsystems, i.e. division by elements. Then we can write

$$E = E_1 + E_2 + U_{12}, V = V_1 + V_2 + V_{12}, N = N_1 + N_2 \qquad (51)$$

where $E_1$ and $E_2$ is the energies of subsystems 1 and 2; $U_{12}$ is the energy of interaction between elements belonging to different subsystems, i.e. the amount of money that elements of different subsystems can simultaneously dispose of; $V_1$ and $V_2$ are the volumes of subsystems 1 and 2, i.e. the amount of goods that belongs only to elements of subsystem 1 or only to elements of subsystem 2; $V_{12}$ is the amount of goods that belongs simultaneously to the elements of subsystem 1 and elements of subsystem 2; $N_1$ and $N_2$ are the number of elements in subsystems 1 and 2.

We assume that the subsystems are economically weakly connected to each other, i.e.

$$|U_{12}| << E_1 + E_2, \ V_{12} \ll V_1 + V_2 \tag{52}$$

In this case, relations (51) take the form

$$E = E_1 + E_2, \ V = V_1 + V_2, N = N_1 + N_2 \tag{53}$$

Let the amount of money that element $i$ can dispose of is $\varepsilon_i$, while the amount of goods belonging to it is equal to $v_i$. Then, in the general case, one can write

$$E = \mathcal{E}(\varepsilon_1, \ldots, \varepsilon_N, v_1, \ldots, v_N), V = \mathcal{V}(\varepsilon_1, \ldots, \varepsilon_N, v_1, \ldots, v_N) \tag{54}$$

Similar dependencies can be written for each of the subsystems of a given system.

In the general case, dependences (54) are nonlinear.

In a particular case, when there is no direct interaction between the elements of the system, i.e. when several elements of the system cannot simultaneously dispose of the same money or the same goods,

$$\mathcal{E} = \varepsilon_1 + \cdots + \varepsilon_N, \mathcal{V} = v_1 + \cdots + v_N \tag{55}$$

The state of the system (market) under consideration is fully characterized by a set of parameters $(\varepsilon_1, \ldots, \varepsilon_N, v_1, \ldots, v_N)$.

The probability of a certain state of the system is $f(\varepsilon_1, \ldots, \varepsilon_N, v_1, \ldots, v_N)d\Omega$, where $d\Omega = d\varepsilon_1 \ldots d\varepsilon_N dv_1 \ldots dv_N$.

By analogy with statistical physics [15], we assume that

$$f(\varepsilon_1, \ldots, \varepsilon_N, v_1, \ldots, v_N) = f(\mathcal{E}(\varepsilon_1, \ldots, \varepsilon_N, v_1, \ldots, v_N), \mathcal{V}(\varepsilon_1, \ldots, \varepsilon_N, v_1, \ldots, v_N))$$

For two weakly interacting systems (markets), i.e. satisfying conditions (52) and (53), we obtain $f(E_1 + E_2, V_1 + V_2, N_1 + N_2) = f(E_1, V_1, N_1)f(E_2, V_2 N_2)$. It follows that the equilibrium economic system is described by a grand canonical distribution

$$f(E, V, N) = Z^{-1} \exp\left(-\frac{E + pV - \mu N}{k_0 T}\right) \tag{56}$$

where $T, p$ and $\mu$ are some parameters characterizing the considered system (market) as a whole; $k_0$ is the numerical constant depending on the choice of unit of measurement for the parameter $T$; $Z$ is the partition function, which is determined from the normalization condition

$$\sum_N \int f(\varepsilon_1, \ldots, \varepsilon_N, v_1, \ldots, v_N) d\Omega = 1 \tag{57}$$

where the summation is performed over all admissible values of the number of elements $N$. Here the parameters $v_i$ (amount of goods) are considered continuous. If the goods are discrete, then integration over $v_i$ must be replaced by summation.

In order for the system to be in equilibrium, certain limitations must be imposed on it; moreover, limitations must be imposed on at least one of the extensive parameters (53).

Consider special cases.

## 5.2. Market with a fixed number of actors

If the number of market actors (elements) $N$ is fixed, then distribution function (56) takes the form

$$f(\varepsilon_1, \ldots, \varepsilon_N, v_1, \ldots, v_N, |N) = Z_N^{-1} \exp\left(-\frac{E+pV}{k_0 T}\right) \tag{58}$$

where the partition function is

$$Z_N(T, p, N) = \int_0^\infty \exp\left(-\frac{\mathcal{E}(\varepsilon_1,\ldots,\varepsilon_N,v_1,\ldots,v_N) + p\mathcal{V}(\varepsilon_1,\ldots,\varepsilon_N,v_1,\ldots,v_N)}{k_0 T}\right) d\Omega \tag{59}$$

Average (for all possible realizations) values of energy and market volume are

$$\overline{E} = \overline{E}(T,p,N) = Z_N^{-1} \int_0^\infty \mathcal{E}(\varepsilon_1, \ldots, \varepsilon_N, v_1, \ldots, v_N) \exp\left(-\frac{\mathcal{E}(\varepsilon_1,\ldots,\varepsilon_N,v_1,\ldots,v_N) + p\mathcal{V}(\varepsilon_1,\ldots,\varepsilon_N,v_1,\ldots,v_N)}{k_0 T}\right) d\Omega \tag{60}$$

$$\overline{V} = \overline{V}(T,p,N) = Z_N^{-1} \int_0^\infty \mathcal{V}(\varepsilon_1, \ldots, \varepsilon_N, v_1, \ldots, v_N) \exp\left(-\frac{\mathcal{E}(\varepsilon_1,\ldots,\varepsilon_N,v_1,\ldots,v_N) + p\mathcal{V}(\varepsilon_1,\ldots,\varepsilon_N,v_1,\ldots,v_N)}{k_0 T}\right) d\Omega \tag{61}$$

As in conventional thermostatistics, knowing the dependence of the partition function (58) on $T, p$ and $N$, one can calculate all the main thermodynamic parameters of the market [15]. In particular, using (59) - (61), we obtain

$$\overline{E} + p\overline{V} = k_0 T^2 \frac{\partial \ln Z_N}{\partial T} \tag{62}$$

$$\overline{V} = -k_0 T \frac{\partial \ln Z_N}{\partial p} \tag{63}$$

We introduce the function

$$G^0(T, p, N) = -k_0 T \ln Z_N \tag{64}$$

Then, taking into account relations (62) - (64), one obtains

$$G^0 = \overline{E} + p\overline{V} + T\left(\frac{\partial G^0}{\partial T}\right)_{p,N} \tag{65}$$

$$\overline{V} = \left(\frac{\partial G^0}{\partial p}\right)_{T,N} \tag{66}$$

Taking into account (66), we write (65) in the form

$$\overline{E} = G^0 - T\left(\frac{\partial G^0}{\partial T}\right)_{p,N} - p\left(\frac{\partial G^0}{\partial p}\right)_{T,N} \qquad (67)$$

One can show that function (64) is the Gibbs free energy of the economic system

$$G^0(T, p, N) = E - TS^0 + pV \qquad (68)$$

Indeed, according to equation (31), function (68) satisfies the equation

$$dG^0 = -S^0 dT + V dp + \mu dN \qquad (69)$$

It follows from equation (69) that

$$S^0 = -\left(\frac{\partial G^0}{\partial T}\right)_{p,N}, V = \left(\frac{\partial G^0}{\partial p}\right)_{T,N}, \mu = \left(\frac{\partial G^0}{\partial N}\right)_{T,p} \qquad (70)$$

In particular, taking into account (18), from the last relation (70) we obtain the well-known thermodynamic relation [15]

$$\mu = \frac{G^0}{N} \qquad (71)$$

Comparing (65) and (66) with (68) and (70), we come to the conclusion that function (64) is indeed the Gibbs free energy (68) of the economic system (market), the parameter $T$ in distribution (58) is the economic temperature of the system, while the parameter $p$ in distribution (58) is the mean price of goods in the market (market pressure).

Thus, relations (60), (61), (66), (67), and (71) determine the average energy, average volume and financial potential of the market at a fixed economic temperature, a fixed price of goods, and a fixed number of market actors.

As in conventional thermostatistics [15], using the partition function (59), one can calculate fluctuations in market parameters, in particular, fluctuations in market energy (amount of money in the market) $\overline{(E - \overline{E})^2} = \overline{E^2} - \overline{E}^2$ and fluctuations of the market volume (amount of goods) $\overline{(V - \overline{V})^2} = \overline{V^2} - \overline{V}^2$ for given values of the parameters $(T, p, N)$. Indeed, using (60) and (61), we obtain

$$k_0 T^2 Z_N^{-1} \frac{\partial Z_N \overline{E}}{\partial T} = \overline{E^2} + p\overline{EV} \qquad (72)$$

$$\overline{V^2} = -k_0 T Z_N^{-1} \frac{\partial Z_N \overline{V}}{\partial p} \qquad (73)$$

$$\overline{EV} = -k_0 T Z_N^{-1} \frac{\partial Z_N \overline{E}}{\partial p} \qquad (74)$$

From (72) and (74) we obtain

$$\overline{E^2} = k_0 T^2 Z_N^{-1} \frac{\partial Z_N \overline{E}}{\partial T} + p k_0 T Z_N^{-1} \frac{\partial Z_N \overline{E}}{\partial p} \qquad (75)$$

Taking into account (64), we write relations (73) and (75) in the form

$$\overline{V^2} = -k_0 T \frac{\partial \overline{V}}{\partial p} + \overline{V} \frac{\partial G^0}{\partial p} \qquad (76)$$

$$\overline{E^2} = k_0 T^2 \frac{\partial \overline{E}}{\partial T} - T\overline{E}\frac{\partial G^0}{\partial T} + G^0 \overline{E} + pk_0 T \frac{\partial \overline{E}}{\partial p} - p\overline{E}\frac{\partial G^0}{\partial p} \tag{77}$$

Taking into account (65), (66), (76) and (77), we obtain relations for fluctuations in market energy and market volume at fixed parameters $(T, p, N)$:

$$\overline{(E - \overline{E})^2} = k_0 T^2 \left(\frac{\partial \overline{E}}{\partial T}\right)_{p,N} + pk_0 T \left(\frac{\partial \overline{E}}{\partial p}\right)_{T,N} \tag{78}$$

$$\overline{(V - \overline{V})^2} = -k_0 T \left(\frac{\partial \overline{V}}{\partial p}\right)_{T,N} \tag{79}$$

### 5.3. Fixed volume market

If the market volume $V$ is fixed, i.e. $\mathcal{V}(\varepsilon_1, \ldots, \varepsilon_N, v_1, \ldots, v_N) = V$, then distribution function (56) takes the form

$$f(\varepsilon_1, \ldots, \varepsilon_N, v_1, \ldots, v_N | V) = Z_V^{-1} \delta_V(\varepsilon_1, \ldots, \varepsilon_N, v_1, \ldots, v_N | V) \exp\left(-\frac{E - \mu N}{k_0 T}\right) \tag{80}$$

where the partition function is

$$Z_V(T, V, \mu) = \sum_{N=0}^{\infty} \int_0^{\infty} \delta_V(\varepsilon_1, \ldots, \varepsilon_N, v_1, \ldots, v_N | V) \exp\left(-\frac{\mathcal{E}(\varepsilon_1, \ldots, \varepsilon_N, v_1, \ldots, v_N) - \mu N}{k_0 T}\right) d\Omega \tag{81}$$

We introduced a generalized δ-function $\delta_V(\varepsilon_1, \ldots, \varepsilon_N, v_1, \ldots, v_N | V)$, which satisfies the conditions

$$\delta_V(\varepsilon_1, \ldots, \varepsilon_N, v_1, \ldots, v_N | V) = \begin{cases} 0, \text{ if } \mathcal{V}(\varepsilon_1, \ldots, \varepsilon_N, v_1, \ldots, v_N) \neq V \\ \infty, \text{ if } \mathcal{V}(\varepsilon_1, \ldots, \varepsilon_N, v_1, \ldots, v_N) = V \end{cases} \tag{82}$$

and

$$\int \delta_V(\varepsilon_1, \ldots, \varepsilon_N, v_1, \ldots, v_N | V) \, dv_1 \ldots dv_N = 1 \tag{83}$$

Average (for all possible realizations) values of energy and number of market elements are

$$\overline{E} = \overline{E}(T, V, \mu) =$$
$$Z_V^{-1} \sum_{N=0}^{\infty} \int_0^{\infty} \mathcal{E}(\varepsilon_1, \ldots, \varepsilon_N, v_1, \ldots, v_N) \delta_V(\varepsilon_1, \ldots, \varepsilon_N, v_1, \ldots, v_N | V) \exp\left(-\frac{\mathcal{E}(\varepsilon_1, \ldots, \varepsilon_N, v_1, \ldots, v_N) - \mu N}{k_0 T}\right) d\Omega \tag{84}$$

$$\overline{N} = \overline{N}(T, V, \mu) = Z_V^{-1} \sum_{N=0}^{\infty} N \int_0^{\infty} \delta_V(\varepsilon_1, \ldots, \varepsilon_N, v_1, \ldots, v_N | V) \exp\left(-\frac{\mathcal{E}(\varepsilon_1, \ldots, \varepsilon_N, v_1, \ldots, v_N) - \mu N}{k_0 T}\right) d\Omega \tag{85}$$

Knowing the dependence of the partition function (81) on $T, V$ and $\mu$, one can find all the main thermodynamic parameters of such a market. In particular, using (81) - (85), we obtain

$$\overline{E} - \mu \overline{N} = k_0 T^2 \frac{\partial \ln Z_V}{\partial T} \tag{86}$$

$$\overline{N} = k_0 T \frac{\partial \ln Z_V}{\partial \mu} \tag{87}$$

We introduce the function

$$\Omega^0(T,V,\mu) = -k_0 T \ln Z_V \tag{88}$$

Then, taking into account relations (62) - (64), we obtain

$$\Omega^0 = \overline{E} - \mu \overline{N} + T\left(\frac{\partial \Omega^0}{\partial T}\right)_{V,\mu} \tag{89}$$

$$\overline{N} = -\left(\frac{\partial \Omega^0}{\partial \mu}\right)_{T,N} \tag{90}$$

Taking into account (90), we write (89) in the form

$$\overline{E} = \Omega^0 - \mu\left(\frac{\partial \Omega^0}{\partial \mu}\right)_{T,N} - T\left(\frac{\partial \Omega^0}{\partial T}\right)_{V,\mu} \tag{91}$$

It is easy to verify that function (88) coincides with the thermodynamic potential of the economic system

$$\Omega^0(T,V,\mu) = E - TS^0 - \mu N \tag{92}$$

Indeed, according to equation (31), thermodynamic potential (92) satisfies the equation

$$d\Omega^0 = -S^0 dT - p dV - N d\mu \tag{93}$$

It follows from equation (93) that

$$S^0 = -\left(\frac{\partial \Omega^0}{\partial T}\right)_{V,\mu}, \quad p = -\left(\frac{\partial \Omega^0}{\partial V}\right)_{T,\mu}, \quad N = -\left(\frac{\partial \Omega^0}{\partial \mu}\right)_{T,V} \tag{94}$$

Taking into account that the parameters $T$ and $\mu$ are intense, and the parameters $\Omega^0$ and $V$ are extensive, we obtain $\Omega^0 = V f_\Omega(T,\mu)$, where $f_\Omega(T,\mu)$ is some function. Then it follows from the second equation (94) that $p = -f_\Omega(T,\mu)$, which is another record of the equation of state (18) of the economic system. As a result, we obtain the well-known thermodynamic relation [15]

$$p = -\frac{\Omega^0}{V} \tag{95}$$

Comparing (89) and (90) with (92) and (94), we come to the conclusion that function (88) is indeed the thermodynamic potential (92) of the economic system (market), the parameter $T$ in distribution (80) is the economic temperature of the system, while the parameter $\mu$ in distribution (80) is the financial potential of the market.

Thus, relations (84), (85), (90), (91), and (95) determine the average energy, the mean price of goods, and the average number of market elements at a fixed economic temperature, a fixed market volume, and a fixed financial potential.

Using the partition function (81), it is easy to calculate fluctuations in market parameters, in particular, fluctuations in market energy (amount of money in the market) and fluctuations in the number of market elements $\overline{(N-\overline{N})^2} = \overline{N^2} - \overline{N}^2$ for given values of parameters $(T,V,\mu)$. Indeed, using (84) and (85), we obtain

$$k_0 T^2 Z_V^{-1} \frac{\partial Z_V \overline{E}}{\partial T} = \overline{E^2} - \mu \overline{EN} \tag{96}$$

$$\overline{N^2} = k_0 T Z_V^{-1} \frac{\partial Z_V \overline{N}}{\partial \mu} \qquad (97)$$

$$\overline{EN} = k_0 T Z_V^{-1} \frac{\partial Z_V \overline{E}}{\partial \mu} \qquad (98)$$

From (96) and (98) we obtain

$$\overline{E^2} = k_0 T^2 Z_V^{-1} \frac{\partial Z_V \overline{E}}{\partial T} + \mu k_0 T Z_V^{-1} \frac{\partial Z_V \overline{E}}{\partial \mu} \qquad (99)$$

Taking into account (88), we write relations (97) and (99) in the form

$$\overline{N^2} = -\overline{N} \frac{\partial \Omega^0}{\partial \mu} + k_0 T \frac{\partial \overline{N}}{\partial \mu} \qquad (100)$$

$$\overline{E^2} = k_0 T^2 \frac{\partial \overline{E}}{\partial T} - T\overline{E} \frac{\partial \Omega^0}{\partial T} + \overline{E} \Omega^0 - \mu \overline{E} \frac{\partial \Omega^0}{\partial \mu} + \mu k_0 T \frac{\partial \overline{E}}{\partial \mu} \qquad (101)$$

Taking into account (90), (91), (100) and (101), we obtain the relations for fluctuations in market energy and number of market elements at fixed parameters $(T, V, \mu)$:

$$\overline{(E - \overline{E})^2} = k_0 T^2 \left(\frac{\partial \overline{E}}{\partial T}\right)_{V,\mu} + \mu k_0 T \left(\frac{\partial \overline{E}}{\partial \mu}\right)_{T,V} \qquad (102)$$

$$\overline{(N - \overline{N})^2} = k_0 T \left(\frac{\partial \overline{N}}{\partial \mu}\right)_{T,V} \qquad (103)$$

### 5.4. Gibbs canonical distribution for market

If both the market volume $V$ and the number of elements $N$ are fixed, then distribution function (56) takes the form

$$f(\varepsilon_1, \ldots, \varepsilon_N, v_1, \ldots, v_N | V, N) = Z_{V,N}^{-1} \delta_V(\varepsilon_1, \ldots, \varepsilon_N, v_1, \ldots, v_N | V) \exp\left(-\frac{E}{k_0 T}\right) \qquad (104)$$

where the partition function is

$$Z_{V,N}(T, V, N) = \int_0^\infty \delta_V(\varepsilon_1, \ldots, \varepsilon_N, v_1, \ldots, v_N | V) \exp\left(-\frac{\mathcal{E}(\varepsilon_1, \ldots, \varepsilon_N, v_1, \ldots, v_N)}{k_0 T}\right) d\Omega \qquad (105)$$

Obviously,

$$Z_N(T, p, N) = \int_0^\infty Z_{V,N}(T, V, N) \exp\left(-\frac{pV}{k_0 T}\right) dV \qquad (106)$$

$$Z_V(T, V, \mu) = \sum_{N=0}^\infty Z_{V,N}(T, V, N) \exp\left(\frac{\mu N}{k_0 T}\right) \qquad (107)$$

The average (for all possible realizations) market energy is

$$\overline{E} = \overline{E}(T, V, N) = Z_{V,N}^{-1} \int_0^\infty \mathcal{E}(\varepsilon_1, \ldots, \varepsilon_N, v_1, \ldots, v_N) \delta_V(\varepsilon_1, \ldots, \varepsilon_N, v_1, \ldots, v_N | V) \exp\left(-\frac{\mathcal{E}(\varepsilon_1, \ldots, \varepsilon_N, v_1, \ldots, v_N)}{k_0 T}\right) d\Omega \qquad (108)$$

Knowing the dependence of the partition function (81) on $T, V$ and $N$, one can calculate all the main thermodynamic parameters of the market. In particular, using (81) - (85), we obtain

$$\overline{E} = k_0 T^2 \frac{\partial \ln Z_{V,N}}{\partial T} \tag{109}$$

We introduce the function

$$F^0(T,V,N) = -k_0 T \ln Z_{V,N} \tag{110}$$

Then, taking into account relations (109) and (110), we obtain

$$F^0 = \overline{E} + T \left(\frac{\partial F^0}{\partial T}\right)_{V,\mu} \tag{111}$$

One can show that, for the economic system under consideration, function (110) is the Helmholtz free energy

$$F^0(T,V,N) = E - TS^0 \tag{112}$$

Taking into account equation (31), we obtain

$$dF^0 = -S^0 dT - p dV + \mu dN \tag{113}$$

Equation (113) implies that

$$S^0 = -\left(\frac{\partial F^0}{\partial T}\right)_{V,N}, p = -\left(\frac{\partial F^0}{\partial V}\right)_{T,N}, \mu = \left(\frac{\partial F^0}{\partial N}\right)_{T,V} \tag{114}$$

Using (111), we calculate

$$d\left(-\frac{\partial F^0}{\partial T}\right) = d\left(\frac{\overline{E} - F^0}{T}\right) = \frac{1}{T}\left[d\overline{E} - \left(\frac{\partial F^0}{\partial T}\right)_{V,N} dT - \left(\frac{\partial F^0}{\partial V}\right)_{T,N} dV - \left(\frac{\partial F^0}{\partial N}\right)_{T,V} dN - (\overline{E} - F^0)\frac{dT}{T}\right] \tag{115}$$

Taking into account (111), we reduce equation (115) to the form

$$T d\left(-\frac{\partial F^0}{\partial T}\right) = d\overline{E} - \left(\frac{\partial F^0}{\partial V}\right)_{T,N} dV - \left(\frac{\partial F^0}{\partial N}\right)_{T,V} dN \tag{116}$$

It is easy to see that equation (31), taking into account (114), coincides with equation (116). Thus, we conclude that the function $F^0(T,V,N)$, defined by relation (110), is the Helmholtz free energy (112) of the economic system (market), the parameter $T$ in distribution (104) is the economic temperature of the system, while the relation (111) is the well-known Gibbs-Helmholtz equation as applied to the economic system (market).

Thus, relations (108), (110), (111), and (114) determine the average energy, the mean price of goods, and the financial potential of the market at a fixed economic temperature, a fixed market volume, and a fixed number of elements.

Using the partition function (81), it is easy to calculate fluctuations in market parameters, in particular, fluctuations in market energy (amount of money in the market) for given values of parameters $(T,V,N)$. Indeed, using (84) and (85), we obtain

$$\overline{E^2} = k_0 T^2 Z_{V,N}^{-1} \frac{\partial Z_{V,N} \overline{E}}{\partial T} \tag{117}$$

Taking into account (110), we write relation (117) in the form

$$\overline{E^2} = k_0 T^2 \frac{\partial \overline{E}}{\partial T} - T\overline{E} \frac{\partial F^0}{\partial T} + \overline{E} F^0 \tag{118}$$

Taking into account (111) and (118), we obtain the relation for market energy fluctuations at fixed parameters $(T, V, N)$:

$$\overline{(E - \overline{E})^2} = k_0 T^2 \left(\frac{\partial \overline{E}}{\partial T}\right)_{V,N} \tag{119}$$

We can calculate the equilibrium fluctuations of the goods price (market pressure) $\overline{(p - \overline{p})^2}$ and the economic temperature of the market $\overline{(T - \overline{T})^2}$, assuming that for fixed $V$ and $N$, they are associated only with energy fluctuations (119).

In this case, using equations of state (11) and (13), we obtain

$$\overline{(p - \overline{p})^2} = \frac{\left(\frac{\partial p}{\partial T}\right)^2_{V,N}}{\left(\frac{\partial E}{\partial T}\right)^2_{V,N}} \overline{(E - \overline{E})^2} \tag{120}$$

$$\overline{(T - \overline{T})^2} = \frac{1}{\left(\frac{\partial E}{\partial T}\right)^2_{V,N}} \overline{(E - \overline{E})^2} \tag{121}$$

Using relation (119) and the well-known thermodynamic relation

$$T \left(\frac{\partial p}{\partial T}\right)_{V,N} = p + \left(\frac{\partial E}{\partial V}\right)_{T,N} \tag{122}$$

which is a consequence of equation (31), for fluctuations (120) and (121) we obtain

$$\frac{\sqrt{\overline{(p-\overline{p})^2}}}{\overline{p}} = \sqrt{\frac{k_0}{\left(\frac{\partial \overline{E}}{\partial T}\right)_{V,N}}} \left(1 + \frac{1}{p}\left(\frac{\partial E}{\partial V}\right)_{T,N}\right) \tag{123}$$

$$\frac{\sqrt{\overline{(T-\overline{T})^2}}}{\overline{T}} = \sqrt{\frac{k_0}{\left(\frac{\partial \overline{E}}{\partial T}\right)_{V,N}}} \tag{124}$$

Similarly, one can calculate the fluctuations in price (pressure), economic temperature and financial potential in other cases discussed above.

### 5.5. Primitive (ideal) market

Thus, economic thermostatistics, like ordinary thermostatistics [15], provides a general method for calculating the thermodynamic parameters of an economic system (market): 1) it is necessary to calculate the partition function $Z$; 2) using the partition function, calculate the corresponding thermodynamic potential (free energy) of the economic system; 3) using the thermodynamic potential and the thermodynamic relations, calculate the average market volume, the average number of system elements, the mean price of goods (market pressure) and the average energy of the system; 4) using the calculated thermodynamic parameters of the market, calculate fluctuations in the price of goods (pressure), energy, volume, number of elements and economic temperature of the market.

Despite the seeming simplicity of this algorithm, the problem of calculating the partition function of the economic system is not at all trivial. In economic thermostatistics, this problem is an order of magnitude more difficult than in conventional thermostatistics [15]. Note that different economic systems (markets) and, accordingly, their models in this approach differ in dependences $\mathcal{E}(\varepsilon_1, \ldots, \varepsilon_N, v_1, \ldots, v_N)$ and $\mathcal{V}(\varepsilon_1, \ldots, \varepsilon_N, v_1, \ldots, v_N)$. Considering that all statistical physics is reduced, in fact, to the development of methods for calculating the partition function, we can expect that we will have a long and interesting work on the development of thermodynamic models of economic systems and methods of approximate calculation of the partition function of various economic systems.

As an example, consider the simplest system (market) consisting of $N$ elements and having a fixed volume of goods $V$. We assume that each unit of goods can be owned (disposed of) by only one market element (actor), i.e.

$$V = v_1 + \cdots + v_N \tag{125}$$

In addition, we assume that the amount of money available in the market (market energy) does not depend on the amount of available goods, i.e.

$$E = \mathcal{E}(\varepsilon_1, \ldots, \varepsilon_N) \tag{126}$$

At the same time, it should be borne in mind that this function does not have to be additive at all, as in (55). Thus, function (126) can be nonlinear if several elements of the system (for example, members of the same family) can dispose of a certain amount of money at the same time.

In the case under consideration, the partition function (105) takes the form

$$Z_{V,N}(T, V, N) = Z_0(T, N) \int_0^\infty \delta_V(v_1, \ldots, v_N | V) dv_1 \ldots dv_N \tag{127}$$

where

$$Z_0(T, N) = \int_0^\infty \exp\left(-\frac{\mathcal{E}(\varepsilon_1, \ldots, \varepsilon_N)}{k_0 T}\right) d\varepsilon_1 \ldots d\varepsilon_N \tag{128}$$

Taking into account that

$$\int_0^\infty \delta_V(v_1, \ldots, v_N | V) dv_1 \ldots dv_N = \int_{v_1 + \cdots + v_{N-1} \leq V} dv_1 \ldots dv_{N-1} = \frac{V^{N-1}}{(N-1)!} \tag{129}$$

and using (110), (114), (127) - (129), one obtains

$$pV = k_0(N - 1)T \tag{130}$$

For $N \gg 1$, relation (130) coincides with the thermal equation of state (10) of the primitive market. So a market that satisfies conditions (125) and (126) is a primitive (ideal) market. Its analog in physics is an ideal gas.

The average energy of the primitive market according to (109) and (127) is

$$\overline{E} = k_0 T^2 \frac{\partial \ln Z_0(T,N)}{\partial T} \tag{131}$$

and does not depend on the volume of the system (the economic analogue of Joule's law for an ideal gas). In the particular case, when each element of the system can independently dispose of a certain amount of money, which belongs only to him, the energy of the system (the amount of money in the system) is equal to the sum of money available to each element of the system (55). In this case, simple calculations (128) give

$$Z_0(T,N) = (k_0 T)^N \tag{132}$$

Substituting (132) into (131), we obtain

$$\overline{E} = k_0 T N \tag{133}$$

Thus, primitive market energy (similar to ideal gas energy) linearly depends on economic temperature. The average amount of money in the primitive market per one element of the system is equal to $\overline{E}/N = k_0 T$, i.e. up to a numerical factor $k_0$ it coincides with the economic temperature of the system. In particular, choosing $k_0 = 1$, we obtain $\overline{E}/N = T$. In this case, the economic temperature is measured in energy units, e.g. in dollars.

Note that in works [3-8], in fact, an ideal (primitive) market with financially independent elements is considered.

Using (123), (124) and (133) for fluctuations in the price of goods and the economic temperature of the primitive market, we obtain

$$\frac{\sqrt{\overline{(p-\overline{p})^2}}}{\overline{p}} = \frac{\sqrt{\overline{(T-\overline{T})^2}}}{\overline{T}} = \frac{1}{\sqrt{N}} \tag{134}$$

Relation (134) is a mathematical expression of the well-known positive role of competition: the higher the competition (the greater the number of elements in the system), the less fluctuations in the price of goods and the more stable the market.

Thermostatistics of the markets will be considered in more detail in the one of the next papers of this series.

## 6. Concluding remarks

In this work, we described a systematic approach to the construction of economic thermodynamics, i.e. the thermodynamic theory of economic systems (markets). The performed analysis shows that using the thermodynamic (thermostatic) approach, it is possible to give economic theory the same rigorous form as theories in physics.

At the same time, when constructing economic thermodynamics, one should keep in mind the important difference between economic systems and physical thermodynamic systems: the ordinary physical thermodynamic systems contain a huge number ($10^{20}$ and more) elements, while the number of elements of economic systems (market actors) rarely exceeds several million; more often it is in the range of $10^4 - 10^5$, and even may be of the order of several hundred. From this point of view, economic systems are closer to physical nanoscale systems. This fact should be taken into account when developing and statistically substantiating economic thermodynamics. In particular, the small size of economic systems can lead to significant fluctuations in their thermodynamic parameters, which makes these systems less stable and less predictable compared to conventional physical thermodynamic systems. At the same time, it is known from physics that even systems consisting of a small number of elements (particles), for example, 100-500, already quite reliably exhibit the thermodynamic properties of real large systems with $10^{20}$ and more elements. This fact underlies such computational methods of statistical physics as the molecular dynamics method and the Monte Carlo method. In particular, the equations of state of real thermodynamic systems can be obtained by calculating systems with a small (up to 1000) number of particles. This gives reason to hope that thermodynamic and thermostatistical methods in economics (with correction to fluctuations) will give acceptable results.

In the next papers of this series, we will show that many well-known economic laws are a natural and formal consequence of economic thermodynamics, and also consider the applications of market thermodynamics to the description of various economic processes.

**Acknowledgments**

This work was done on the theme of the State Task No. AAAA-A20-120011690135-5. Funding was provided in part by the Tomsk State University competitiveness improvement program.